# Comparison of Different Methods for Calculation of the Normalized Mott Cross Section[1]

P.B. Kats*[2], K.V. Halenka*, O.O. Voskresenskaya**[3]

*Brest State A.S. Pushkin University, Brest, 224016, Belarus
**Joint Institute for Nuclear Research, Dubna, Moscow Region, 141980 Russia

*An intercomparison of some earlier methods for calculating the normalized Mott cross section and also a method proposed by the authors of the present work is carried out. It is demonstrated that applying the given method, along with the method of Lijian et al., is preferable for relevant calculations.*

## 1. Introduction

Knowledge of the energy loss of ions in matter, commonly described as stopping power (the ion mean energy loss per unit travelled path length), is fundamental to many applications dependent on the transport of ions in matter, particularly ion implantation, ion-beam modification of materials, ion beam-analysis, and ion-beam therapy [1−3].

The electronic stopping power of a material is described by the Bethe formula (the so-called Bethe's stopping power formula) [4, 5]. Its relativistic version was obtained by him in [5]. Taking into account the density effect, the average ionization energy loss by moderately relativistic charged heavy particles can be described in the first Born approximation as follows:

$$-\frac{d\bar{E}}{dx} = \zeta L, \quad L = L_0 = \ln\left(\frac{E_m}{I}\right) - \beta^2 - \frac{\delta}{2},$$

$$\zeta = 4\pi r^2 mc^2 \cdot N_e \cdot \left(\frac{Z}{\beta}\right)^2 = 4\pi r^2 mc^2 \cdot N_A \rho \frac{Z'}{A} \cdot \left(\frac{Z}{\beta}\right)^2, \quad E_m = \frac{2mc^2\beta^2}{1-\beta^2}. \quad (1)$$

The function $L_0$ of this form is derived originally from quantum perturbation theory. The first two of its terms are typically called the Bethe result and the third term is the familiar density effect correction of Fermi [6, 7]. In these equations, $x$ denotes the distance traveled by a particle; $E_m$ is the maximum transferrable energy to an electron of mass $m$ and classical radius $r$ in a collision with the particle of velocity $\beta c$; $I$ is the effective ionization potential of the absorber atoms, $Z$ is the charge number of incident nucleus, and $N_e$ is the electron density of a material. The electron density is either measured in electrons/g ($\tilde{N}_e = N_A Z'/A$) or in electrons/cm$^3$ ($N_e = N_A \rho Z'/A$), where $\rho$ is density of a material in g cm$^{-3}$, $N_A$ denotes the Avogadro number, $Z'$ and $A$ refer to the atomic number and weight of the absorber. In the latter case, we can rewrite Eq. (1) as

---

$$-\frac{d\bar{E}}{\rho dx} = \tilde{\zeta}L, \quad \tilde{\zeta} = 4\pi r^2 mc^2 \cdot \tilde{N}_e \cdot \left(\frac{Z}{\beta}\right)^2 = 0.307075 \frac{Z'}{A}\left(\frac{Z}{\beta}\right)^2.$$

This is the 'mass stopping power' in units MeV g$^{-1}$cm$^2$ [8].

The above expressions are applicable if $Z\alpha/\beta \ll 1$, where $\alpha$ is the fine-structure constant. If this condition is not satisfied, the Bloch corrections $\Delta L_B$ [9] and the Mott corrections $\Delta L_M$ [10, 11] are also introduced:

$$\Delta L_B = \psi(1) - \operatorname{Re}\psi(1+iZ\alpha/\beta)$$

with the digamma function $\psi$ and

$$\Delta L_M = \frac{\tilde{N}_e}{\tilde{\zeta}} \int_\varepsilon^{E_m} E\left[\left(\frac{d\sigma}{dE}\right)_M - \left(\frac{d\sigma}{dE}\right)_B\right]dE. \tag{2}$$

Here, $\varepsilon$ is some energy above which the atomic electron binding energy may be neglected, and $(d\sigma/dE)_{M(B)}$ are, respectively, the Mott and Born expressions for the scattering cross section of electrons on nuclei. Switching in the expression (2) from integration over the energy $E$ transferred to an electron to integration over the center-of-mass scattering angle $\theta$, we can rewrite (2) in the form

$$\Delta L_M = 2\pi \frac{\tilde{N}_e E_m}{\tilde{\zeta}} \int_{\theta_0}^{\pi}\left[\left(\frac{d\sigma(\theta)}{d\Omega}\right)_M - \left(\frac{d\sigma(\theta)}{d\Omega}\right)_B\right]\sin^2\left(\frac{\theta}{2}\right)\sin\theta d\theta, \tag{3}$$

where $\theta_0$ denotes the scattering angle corresponding to $\varepsilon$ and $\Omega$ is the usual scattering cross section solid angle.

The Mott correction was first calculated by Eby and Morgan [12, 13] by numerical integration of (2) for several values of $Z$ and $\beta$. These calculation demonstrated the importance of taking account of Mott's corrections to the Bethe−Bloch formula for incident nuclei with $Z \geq 20$. Since the expressions (2), (3) for $\Delta L_M$ are extremely inconvenient for practical application, the analytical expressions for $\Delta L_M$ in the second and third order Born approximations were also proposed in [13]. Significant simplification of computing the Mott corrections is provided by a method of [14] that reduces the problem to the numerical summation of an infinite series.

This paper presents an adaptation of the method [14] for calculating the Mott differential cross section (MDCS) normalized with respect to the Rutherford differential cross section (RDCS), as well as a comparison of this adopted method with some other rigorous and approximate methods for relevant calculations. The communication is organized as follows. Section 2 considers some preliminaries that used later in Section 3, i.e. a standard description of the (normalized) MDCS (Section 2.1) and the different approximations to the normalized Mott cross section (Section 2.2). Section 3 presents an another exact representation for the normalized MDCS (Section 3.1) and an intercomparison of applying all the mentioned methods (Section 3.2). Section 4 contains a summary of our results and conclusions.

## 2. Preliminaries

### 2.1 Mott's differential cross section

In 1911 Rutherford calculated the differential cross section for scattering of electrons by the Coulomb potential in the framework of classical mechanics [15], obtaining the well-known Rutherford formula:

$$\sigma_R \equiv \left(\frac{d\sigma}{d\Omega}\right)_R = \left(\frac{Ze^2}{2mv^2}\right)^2 \frac{1}{\sin^4(\theta/2)}. \tag{4}$$

Within the framework of nonrelativistic quantum mechanics, a solution to this problem was found independently by Gordon [16] and Mott [17] in 1928. Six months later, a simpler solution was proposed by Temple [18].

An expression for the scattering cross section of relativistic electrons through the Coulomb potential (Eq. 5) was provided by Mott in 1929–1932 [10, 11]. This expression cannot be given in analytical form and contains slowly converging infinite series of Legendre polynomials ($P_k$):

$$\sigma_M \equiv \left(\frac{d\sigma}{d\Omega}\right)_M = \left(\frac{\hbar}{mv}\right)^2 (1-\beta^2)\left(\frac{\xi^2 |F_M|^2}{\sin^2(\theta/2)} + \frac{|G_M|^2}{\cos^2(\theta/2)}\right), \tag{5}$$

where

$$F_M(\theta) = \frac{1}{2}i\sum_{k=0}^{\infty}(-1)^k[kC_M^{(k)} + (k+1)C_M^{(k+1)}]P_k(\cos\theta) = \sum_{k=0}^{\infty}F_M^{(k)}P_k(\cos\theta),$$

$$G_M(\theta) = \frac{1}{2}i\sum_{k=0}^{\infty}(-1)^k[k^2C_M^{(k)} - (k+1)^2C_M^{(k+1)}]P_k(\cos\theta) = \sum_{k=0}^{\infty}G_M^{(k)}P_k(\cos\theta),$$

with

$$C_M^{(k)} = -e^{-i\pi\rho_k}\frac{\Gamma(\rho_k - i\eta)}{\Gamma(\rho_k + 1 + i\eta)}, \quad \eta = \frac{Z\alpha}{\beta}, \quad \xi = \eta\sqrt{1-\beta^2}, \quad \rho_k = \sqrt{k^2 - (Z\alpha)^2}, \quad \alpha = \frac{e^2}{\hbar c}.$$

Here, $F_M(\theta)$ and $G_M(\theta)$ are two complex functions,

$$F_M(\theta) = F_0(\theta) + F_1(\theta), \quad G_M(\theta) = G_0(\theta) + G_1(\theta), \tag{6}$$

with

$$F_0(\theta) = \frac{1}{2}i\sum_{k=0}^{\infty}(-1)^k[kC_Z^{(k)} + (k+1)C_Z^{(k+1)}]P_k(\cos\theta),$$

$$G_0(\theta) = \frac{1}{2}i\sum_{k=0}^{\infty}(-1)^k[k^2 C_Z^{(k)} - (k+1)^2 C_Z^{(k+1)}]P_k(\cos\theta),$$

$$F_1(\theta) = \frac{1}{2}i\sum_{k=0}^{\infty}(-1)^k[kD^{(k)} + (k+1)D^{(k+1)}]P_k(\cos\theta),$$

$$G_1(\theta) = \frac{1}{2}i\sum_{k=0}^{\infty}(-1)^k[k^2 D^{(k)} - (k+1)^2 D^{(k+1)}]P_k(\cos\theta),$$

where the functions $C_Z^{(k)}$ and $D^{(k)}$ are as follows:

$$C_Z^{(k)} = -e^{-i\pi k}\frac{\Gamma(k-i\eta)}{\Gamma(k+1+i\eta)}, \quad D^{(k)} = C_M^{(k)} - C_Z^{(k)}.$$

Hence, the functions $F_0(\theta)$ and $G_0(\theta)$ may be written as

$$F_0(\theta) = \frac{i}{2}\frac{\Gamma(1-i\eta)}{\Gamma(1+i\eta)}\exp\left\{i\eta\ln\left[\sin^2\left(\frac{\theta}{2}\right)\right]\right\}, \quad G_0(\theta) = -i\eta\frac{F_0(\theta)}{\tan^2(\theta/2)}. \quad (7)$$

The formula (5) is also referred to as an exact formula for the differential cross section, because no Born approximation of any order is used in its derivation.

The first numerical summation of above series was performed by Mott himself [11] for scattering of electrons with relative velocity $\beta$ from 0.1 to 1.0 by gold nuclei (Z = 79) at 90 degrees. Starting from this work, in such calculations began to introduce a quantity equal to the ratio of the MDCS ($\sigma_M$) to the modified RDCS ($\tilde{\sigma}_R$),

$$R(\theta) = \sigma_M/\tilde{\sigma}_R, \quad \tilde{\sigma}_R = \sigma_R(1-\beta^2), \quad (8)$$

i.e., the normalized Mott cross section (NMCS). In ref. [20], the indicated quantity has the form:

$$R_M(\theta) = \frac{4\sin^2(\theta/2)}{\eta^2}\left[\xi^2|F_M|^2 + \tan^2\left(\frac{\theta}{2}\right)|G_M|^2\right]. \quad (9)$$

Since the 'exact' MDCS (5) and NMCS (9) are expressed in terms of slowly converging Legendre polynomial series, their application to calculate integrals (2), (3) is a difficult problem. In this regard, the use of analytical approximations to them and getting other their representations becomes important.

## 2.2 Some approximations to the normalized Mott differential cross section

One way to obtain such approximations is to expand the exact NMCS in terms of power series in $\alpha Z$. We will present below such results for the above function $R(\theta)$.

The first such expansion was obtained by the author of the exact solution to the scattering problem [11]:

$$R_B(\theta) = 1 - \beta^2 \sin^2\left(\frac{\theta}{2}\right). \qquad (10)$$

Further approximations were obtained by McKinley and Feshbach,

$$R_{MF}(\theta) = R_B + \alpha Z \pi \beta \sin\left(\frac{\theta}{2}\right)\left[1 - \sin\left(\frac{\theta}{2}\right)\right], \qquad (11)$$

as well as Johnson, Weber, and Mullin [19, 20],

$$\begin{aligned}R_{JWM} &= R_{MF} + (\alpha Z)^2 \sin\left(\frac{\theta}{2}\right)\bigg\{ L_2\left[1-\sin^2\left(\frac{\theta}{2}\right)\right] - 4L_2\left[1-\sin\left(\frac{\theta}{2}\right)\right] + 2\sin\left(\frac{\theta}{2}\right)\ln^2\left[1-\sin\left(\frac{\theta}{2}\right)\right] \\ &+ \frac{\pi^2}{2}\left[1-\sin\left(\frac{\theta}{2}\right)\right] + \frac{\pi^2}{6}\sin\left(\frac{\theta}{2}\right) + \beta^2 \sin\left(\frac{\theta}{2}\right)\left(L_2\left[1-\sin^2\left(\frac{\theta}{2}\right)\right] + \frac{\sin^2(\theta/2)\ln^2[\sin(\theta/2)]}{1-\sin^2(\theta/2)}\right) \\ &+ \frac{\pi^2}{4}\frac{1-\sin(\theta/2)}{1+\sin(\theta/2)} - \frac{\pi^2}{6}\bigg\},\end{aligned} \qquad (12)$$

where $L_2$ denotes Euler's dilogarithm defined by

$$L_2(x) = \int_0^x \frac{\ln(1-y)}{y} dy.$$

Another approach was proposed by Lijian, Qing, and Zhengming [21], where the exact NMCS is approximated by the following expression:

$$R_{LQZ}(\theta; Z, E) = \sum_{j=0}^{4} a_j(Z,E)(1-\cos\theta)^{j/2}, \quad a_j(Z,E) = \sum_{k=1}^{6} d_Z(j,k)(\beta - \bar{\beta})^{k-1}, \quad \bar{\beta} = 0.7181287. \qquad (13)$$

The authors calculated 30 coefficients $d_Z(j,k)$ for 90 elements of the Periodic System with target atomic number $Z$ from 1 to 90 in a wide range of energy. Investigations in this direction were continued by Boschini, Consolandi, Gervasi et al. in the work [22], where the coefficients $d_Z(j,k)$ were obtained for 118 elements of the Periodic Table of Elements both for electrons and positrons.

# 3. Results and discussions

## 3.1 Another representation for the normalized Mott cross section

In [14] we got the following representation for the exact Mott differential cross section:

$$\sigma_{VSST} \equiv \left(\frac{d\sigma}{d\Omega}\right)_{VSTT} = \left(\frac{\hbar}{mv}\right)^2 (1-\beta^2)\left(\frac{\xi^2 |F_M|^2 - |F'_M|^2}{\sin^2(\theta/2)}\right) \equiv \left(\frac{\hbar}{mv}\right)^2 (1-\beta^2)\omega_{VSST},$$

$$\omega_{VSTT}(\theta) = \omega_Z(\theta) + \lambda(\theta)/\sin^2\left(\frac{\theta}{2}\right),$$

$$\lambda(\theta)/4 = \xi^2\left[2\operatorname{Re}(\Delta F F_Z^*) + |\Delta F|^2\right] + 2\operatorname{Re}(\Delta F' F_Z^{*\prime}) + |\Delta F'|^2,$$

$$\omega_Z(\theta) = \left[\xi^2 + \eta^2 \cos^2\left(\frac{\theta}{2}\right)\right]/\sin^2\left(\frac{\theta}{2}\right) \equiv \omega_B(\theta), \quad \Delta F \equiv F_M - F_Z,$$

$$F_Z(\theta) = \frac{i}{2}\sum_{l=0}^{\infty}(-1)^k F_Z^{(k)} P_k(\cos\theta) = \frac{i}{2}\frac{\Gamma(1-i\eta)}{\Gamma(1+i\eta)}\sin^{2i\eta}\left(\frac{\theta}{2}\right), \quad F_Z^{(k)} = kC_Z^{(k)} + (k+1)C_Z^{(k+1)},$$

$$F'_M \equiv dF_M(\theta)/d\theta = -\tan^2(\theta/2)G_M. \qquad (14)$$

This representation reduces computing the integrals (2), (3) to a summing the fast converging infinite series whose terms are bilinear in the Mott partial amplitudes and can be simply implemented using the numerical summation methods of converging series for a given level of precision.

It leads to the following exact expression for the normalized Mott cross section (see Appendix):

$$R_{KHV}(\theta) = R_B(\theta) + \tilde{\lambda}(\theta)\sin^2\left(\frac{\theta}{2}\right),$$

$$\tilde{\lambda}(\theta)/4 = \eta^{-2}\left\{\xi^2\left[2\operatorname{Re}(\Delta F F_Z^*) + |\Delta F|^2\right] + 2\operatorname{Re}(\Delta F' F_Z^{*\prime}) + |\Delta F'|^2\right\}. \qquad (15)$$

Taking into account (6), (7), we can rewrite $\tilde{\lambda}(\theta)$ in terms of functions $F_0(\theta)$ and $F_1(\theta)$,

$$\tilde{\lambda}(\theta)/4 = \eta^{-2}\left\{\xi^2\left[2\operatorname{Re}(F_1 F_0^*) + |F_1|^2\right] + 2\operatorname{Re}(F_1' F_0^{*\prime}) + |F_1'|^2\right\},$$

and then calculate the ratio $R_{KHV}(\theta)$, for instance, by Sherman's method [23].

## 3.2 Comparison of methods

Table 1 lists the results of calculating the normalized Mott cross section $R(\theta)$ by the above methods. It shows an excellent agreement between the results obtained from Eqs. (15) and (9) as well as an increasing deviation from these results in the transition from (13) to (10). This allows us to carry out further comparison with respect to the results obtained on the basis of (15).

**Table 1**: Comparison of the $R(\theta)$ values obtained by different methods for the scattering of electrons with an energy of 10 MeV on nuclei of charge number $Z = 47$.

| $R/\theta$ | 15 | 30 | 45 | 60 | 75 | 90 | 105 | 120 | 135 | 150 | 165 | 180 |
|---|---|---|---|---|---|---|---|---|---|---|---|---|
| $R_M$ | 1.116 | 1.215 | 1.256 | 1.226 | 1.122 | 0.958 | 0.753 | 0.533 | 0.324 | 0.154 | 0.042 | 0.0032 |
| $R_{KHV}$ | 1.116 | 1.215 | 1.256 | 1.226 | 1.122 | 0.958 | 0.753 | 0.533 | 0.324 | 0.154 | 0.042 | 0.0032 |
| $R_{LQZ}$ | 1.118 | 1.214 | 1.255 | 1.225 | 1.123 | 0.959 | 0.753 | 0.532 | 0.323 | 0.153 | 0.043 | 0.0041 |
| $R_{JWM}$ | 1.143 | 1.228 | 1.240 | 1.171 | 1.042 | 0.867 | 0.667 | 0.463 | 0.278 | 0.131 | 0.036 | 0.0032 |
| $R_{MF}$ | 1.105 | 1.140 | 1.108 | 1.020 | 0.886 | 0.724 | 0.549 | 0.377 | 0.224 | 0.105 | 0.029 | 0.0026 |
| $R_B$ | 0.983 | 0.933 | 0.854 | 0.751 | 0.630 | 0.501 | 0.372 | 0.252 | 0.149 | 0.069 | 0.019 | 0.0026 |

Figure 1 compares the results obtained on the basis of Eqs. (10)−(13), (15) for scattering of electrons with energies of 0.005 MeV, 1 MeV, and 10 MeV on nuclei of charge number $Z = 13$, 47, and 92.

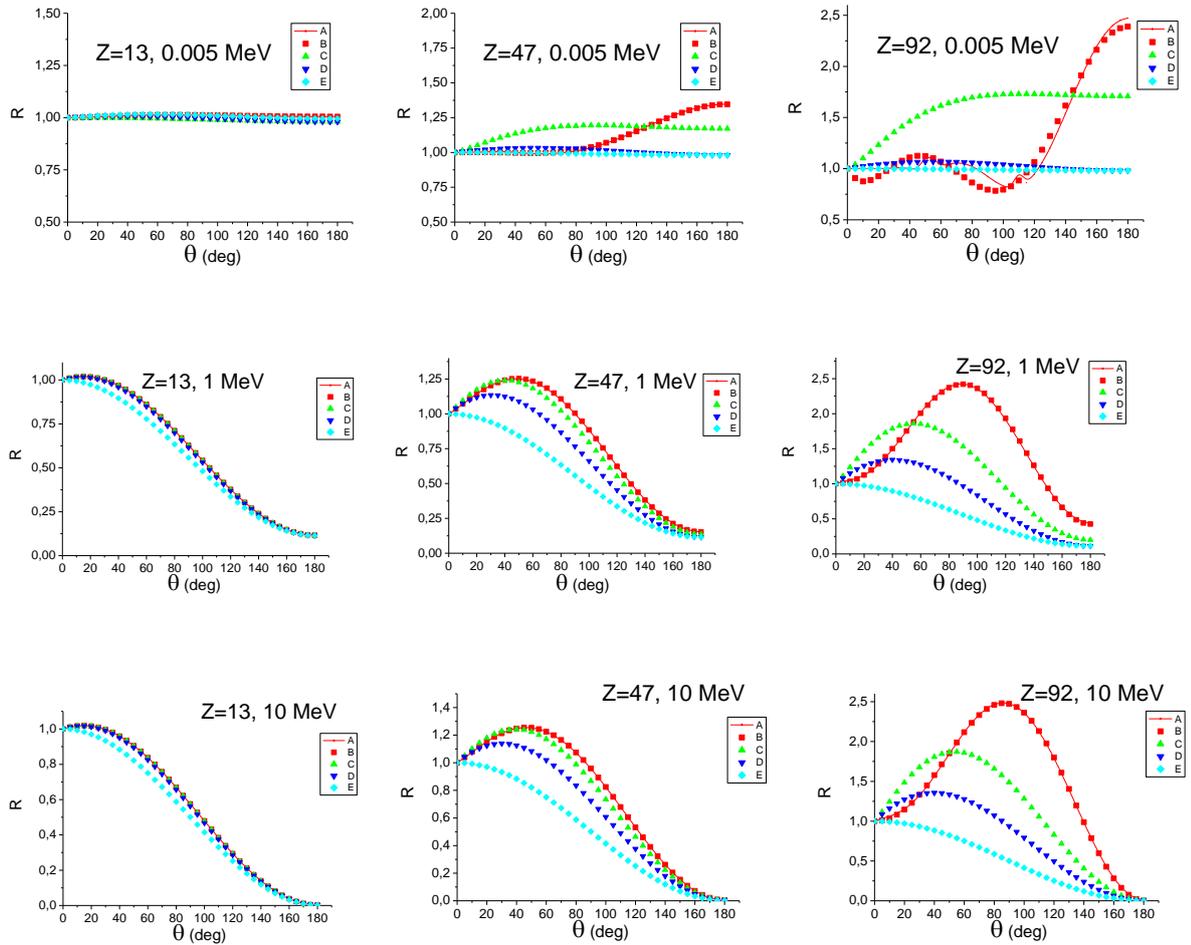

Figure 1: Cross section ratio, $R(\theta)$, as function of scattering angle obtained from Eqs. 15 (A), 13 (B), 12 (C), 11 (D), 10 (E) for scattering of electrons with energies of 0.005 MeV, 1 MeV, and 10 MeV on nuclei of charge number $Z$ equal to 13, 47, and 92.

From this Figure it can be seen that the results of Lijian et al. and Boschini et al. [21, 22] obtained from Eq. (13) significantly differ from the exact ones only in the area of low energies and high charge numbers (e.q. for $Z = 92$, 0.005 MeV). In other cases, they are close to rigorous results. For light elements, all approximations give fairly accurate results. For elements with moderately high values of $Z$ at medium and high energies, the approximation (12) gives higher accuracy than (11) and (10). For heavy elements, the approximate methods based on Eqs. (10)−(12) are not applicable.

Additionally we evaluated relative difference between the ratios $R_{LQZ}$ and $R_{KNV}$ obtained by the methods of works [21, 22] and [14] as a function of the scattering angle for electrons with energies from 0.005 MeV to 10 MeV on nuclei with charge number from 13 to 92 (Figure 2):

$$\delta R(\theta; Z, E) = \frac{R_{LQZ}(\theta; Z, E) - R_{KHV}(\theta; Z, E)}{R_{KHV}(\theta; Z, E)}.$$

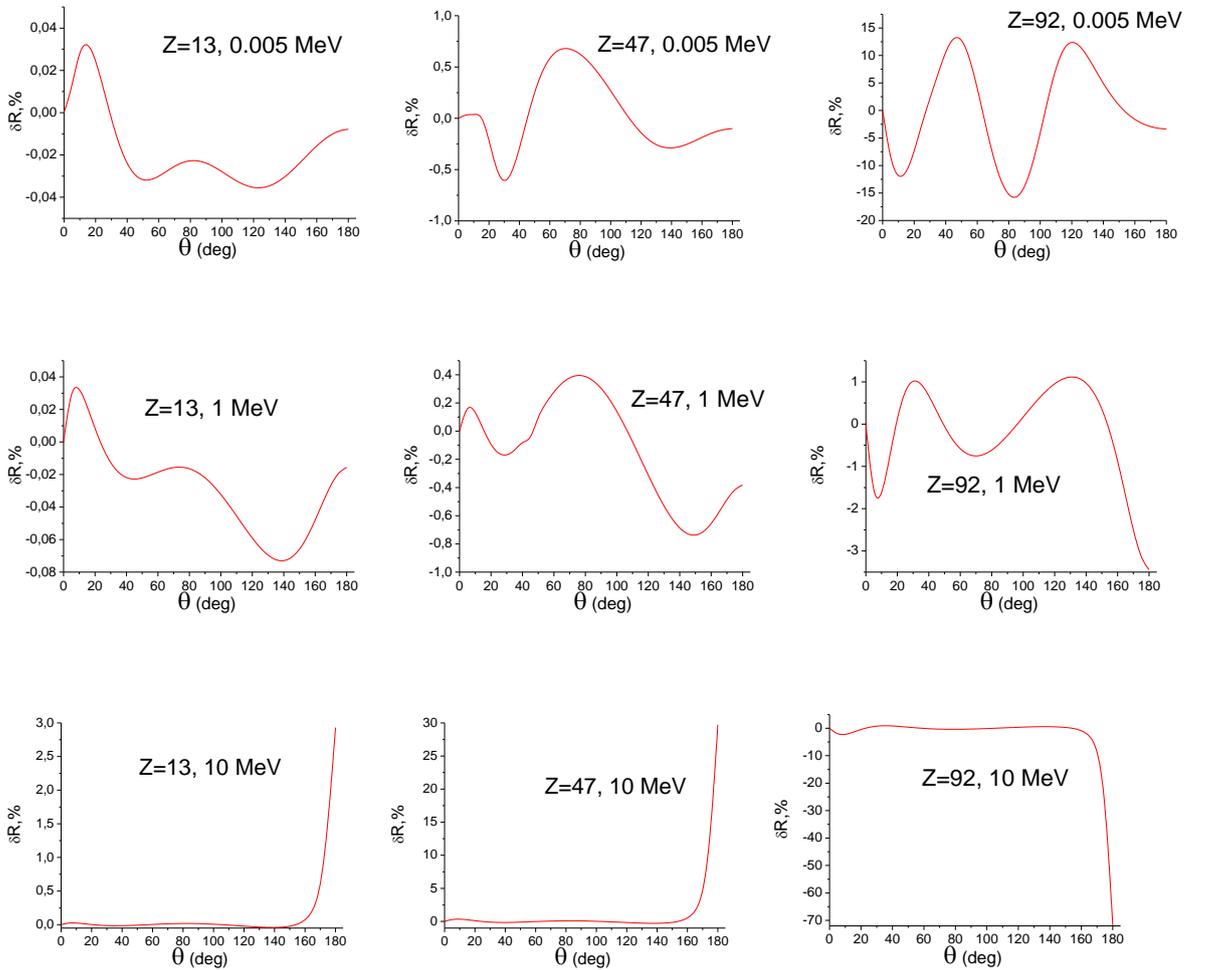

Figure 2: Relative difference between the ratios $R_{LQZ}$ and $R_{KHV}$ obtained from Eqs. (13) and (15) as function of scattering angle (in degrees) for electrons with energies of 0.005 MeV, 1 MeV, and 10 MeV scattered on nuclei of charge number $Z = 13, 47,$ and 92.

Figure 2 shows that at low energies (e.g. 0.005 MeV), the maximum value of the relative difference modulus $|\delta R(\theta; Z, E)|$ increases from 0.003 to 16 percent in the transition from nucleus charge number $Z = 13$ to $Z = 92$. From Figure 2 also follows that at medium energies (1 MeV), this value varies between 0.07−3.5 percent for nucleus with a $Z$ value of 13 to 92. At high energies[4] (e. q. 10 MeV), the approximation (13) differs significantly (up to 70 percent) from the exact expression (15) only in the range of scattering angles from 160 to 180 degrees, where the values of the ratios $R_{LQZ}$ and $R_{KHV}$ are very small, while over the $\theta$ range from 0 to 150 degrees, the relative difference between $R_{LQZ}$ and $R_{KHV}$ is almost zero.

## 4. Summary and conclusions

- In the present work, an new exact representation for the normalized MDCS is proposed that reduces the calculation of the NMCS in terms of the Mott series $F_M(\theta)$ and $G_M(\theta)$ to its calculation in terms of $F_M(\theta)$ alone, excluding the most slowly converging series in the NMCS computation.

- Numerical results are obtained on the basis of the obtained formula and the following exact and approximate expressions for the normalized Mott cross section: i) the conventional Mott-exact 'phase-shift' formula (point-charge nucleus, no screening) [11], ii) the approximate Lijian−Qing−Zhengming expression [21], iii) the Johnson−Weber−Mullin formula [20], iv) the McKinley−Feshbach expression [19], and v) the Mott−Born result [11].

- An intercomparison of the obtained numerical results is presented in the range of nucleus charge number from $Z = 13$ to $Z = 92$ for electron energies from 0.005 MeV to 10 MeV and scattering angles over the range of 0−180 degrees.
    - It is shown that while all the methods discussed give sufficiently accurate results for low-Z nuclei in the entire range of energies, the approximate Mott−Born, McKinley−Feshbach, and Johnson−Weber− Mullin methods are not applicable for high-Z nuclei at the same energies.
    - The approximate Lijian−Qing−Zhengming method gives fairly accurate results in the entire range of charge numbers and electron energies, except for the area of low energies and high charge numbers.
    - The results of the rigorous methods considered are remarkably consistent.

---

[4] At energies higher than 10 MeV, the results are very close to those of 10 MeV, according to [21], since $\beta$ in this case is close to 1.

- The accuracy was estimated, and the range of applicability was established for the Lijian−Qing−Zhengming method, which gives the best approximation to rigorous results.
    - We managed to show that for $Z < 90$, this method can be applied with an error of less than 1%, in accordance with [21], but only over the $\theta$ range from 0 to 150 degrees at high energies.
    - In the case of $Z \geq 90$, the specified method can also be applied with the same error, however also only in the $\theta$ range of 0−150 degrees for high and medium energies.
    - Outside of the specified ranges, the error can increase up to 16 percent (for $Z = 92$, 0.005 MeV) and even up to 70% (for $Z = 92$, 10 MeV, and $\theta = 180$ degrees).
- Thus, we can conclude that both the rigorous method suggested in this work and the approximate Lijian−Qing−Zhengming method can be recommended for practical calculations of the normalized Mott cross section $R(\theta)$.
    - Although the second method has somewhat limited accuracy, its advantage compared to first method is the ability to perform integration with a given lower integration limit.
    - The advantage of the first method over the second one is its greater accuracy, as well as the possibility of its use beyond the applicability of the approximate method by Lijian, Qing, and Zhengming.
    - Therefore, each of these methods is preferred in its application area for relevant calculations of the NMCS.

## Appendix: Derivation of the formula for the normalized Mott cross section

The derivation of the formula (15) in terms of $F_0(\theta)$ and $F_1(\theta)$ can be represented as follows.

Substituting the expression for $G_M$ from (14) into (9) and taking into account (6), we have

$$R_M(\theta) = \frac{4\sin^2(\theta/2)}{\eta^2}\left[\xi^2 |F_M|^2 + |F_M'|^2\right] = \frac{4\sin^2(\theta/2)}{\eta^2}\left[\xi^2 |F_0 + F_1|^2 + |F_0' + F_1'|^2\right]. \quad (16)$$

After carrying out a number of transformations in (16), we can write it as

$$R_M(\theta) = \frac{4\sin^2(\theta/2)}{\eta^2}\left[\xi^2 (F_0 + F_1)(F_0^* + F_1^*) + (F_0' + F_1')(F_0'^* + F_1'^*)\right] =$$
$$= \frac{4\sin^2(\theta/2)}{\eta^2}\left\{\xi^2 |F_0|^2 + |F_0'|^2 + \xi^2\left[2\operatorname{Re}(F_1 F_0^*) + |F_1|^2\right] + 2\operatorname{Re}(F_1' F_0'^*) + |F_1'|^2\right\},$$

$$|F_0|^2 = \frac{1}{4}, \quad |F_0'|^2 = \frac{\eta^2}{4\tan^2(\theta/2)}.$$

Thus, we get as a result the following expression for the normalized Mott cross section:

$$R_{KHV}(\theta) = R_B(\theta) + \tilde{\lambda}(\theta)\sin^2\left(\frac{\theta}{2}\right),$$

$$\tilde{\lambda}(\theta) = \frac{4}{\eta^2}\left\{\xi^2\left[2\operatorname{Re}(F_1 F_0^*) + |F_1|^2\right] + 2\operatorname{Re}(F_1' F_0^{*\prime}) + |F_1'|^2\right\},$$

$$R_Z(\theta) = \frac{4\sin^2(\theta/2)}{\eta^2}\left[\frac{\xi^2}{4} + \frac{\eta^2}{4\tan^2(\theta/2)}\right] = 1 - \beta^2\sin^2(\theta/2) \equiv R_B(\theta). \tag{17}$$

This normalized Mott cross section (17) can be calculated for example by Sherman's method. However, our calculations show that already elimination from (5) of the slowest converging function $G_1(\theta)$ provides a convergence of these series comparable to that obtained by a 'method of reduced series' used in his work [23].